# Non-Orthogonal Multiple Access (NOMA): How It Meets 5G and Beyond

S. M. Riazul Islam, Ming Zeng, Octavia A. Dobre, and Kyung-Sup Kwak

S.M. Riazul Islam, Department of Computer Science and Engineering, Sejong University, Seoul, South Korea; riaz@sejong.ac.kr

Ming Zeng, Faculty of Engineering and Applied Science, Memorial University, Canada; mzeng@mun.ca

Octavia A. Dobre, Faculty of Engineering and Applied Science, Memorial University, Canada; odobre@mun.ca

Kyung-Sup Kwak, School of Information and Communication Engineering, Inha University, South Korea; kskwak@inha.ac.kr

**Abstract:** Due to massive connectivity and increasing demands of various services and data-hungry applications, a full-scale implementation of the fifth generation (5G) wireless systems requires more effective radio access techniques. In this regard, non-orthogonal multiple access (NOMA) has recently gained ever-growing attention from both academia and industry. Compared to orthogonal multiple access (OMA) techniques, NOMA is superior in terms of spectral efficiency and is thus appropriate for 5G and Beyond. In this article, we provide an overview of NOMA principles and applications. Specifically, the article discusses the fundamentals of power-domain NOMA with single and multiple antennas in both uplink and downlink settings. In addition, the basic principles of code-domain NOMA are elaborated. Further, the article explains various resource allocation techniques such as user pairing and power allocation for NOMA systems; discusses the basic form of cooperative NOMA and its variants; and addresses several opportunities and challenges associated with the compatibility of NOMA with other advanced communication paradigms such as heterogeneous networks and millimeter wave communications.

# 1. Introduction

Every generation of cellular networks comes with new standards, techniques and features, differentiating it from the previous one. In line with that, the next generation cellular systems, the fifth generation (5G) and beyond, are expected to support various advanced services including multimedia applications, Internet-of-Things (IoT)-based applications, and vehicle-to-everything (V2X) [1]. These innovative use cases are leading the gigantic growth of mobile traffic, which is in turn introducing radio spectrum scarcity as one of the most critical challenges that 5G and Beyond should deal with.

Multiple access, one of the fundamental building blocks in wireless communication systems, has a significant impact on the utilization of the available spectrum, system throughput and latency. In cellular radio context, multiple access refers to a technique by which multiple users share a common radio resource to establish communication links with a base station (BS). Some of the widely used multiple access techniques in the past generations of cellular networks include time division multiple access (TDMA), frequency division multiple access (FDMA), and code division multiple access (CDMA). These techniques are referred to as orthogonal multiple access (OMA); the access of users is orthogonal in nature and, ideally, the users do not interfere with one another while they share the communication channel. In these schemes, orthogonal radio resources in time-, frequency-, code-domain or their combinations are assigned to multiple users. In TDMA, a dedicated time slot is assigned to each user to transmit their signals. At the receiving ends, users' signals are thus differentiated by identifying the corresponding time slots. Similarly, FDMA divides the available bandwidth (BW) into a number of non-overlapping frequency subchannels, allocates each subchannel to a separate user, and allows multiple users to send their signals using the assigned subchanels. At the receiving ends, users' signals are thus differentiated by identifying the respective frequency bands. Unlike FDMA and TDMA, CDMA assigns a unique orthogonal spreading sequence to each user and users transmit their signals sharing the same time-frequency resources, each using its allocated spreading sequence. At the receiving ends, users' signals are separated by employing decorrelation that identifies the desired user's signal based on the respective assigned code, while treating the other users' signals as noise.

Theoretically, OMA-based systems do not experience inter-user interferences because of orthogonal resource allocation, and therefore, typically low-complexity receivers can be employed to detect the signal of the desired user. However, as the number of orthogonal resources is limited, the OMA systems cannot serve a large number of users, as imposed by 5G. In contrast to OMA, non-orthogonal multiple access (NOMA) allows inter-user interference in the resource allocation of users and thus multiple users are served using the same resource block. To mitigate the effect of the interference, interference cancellation schemes such as successive interference cancellation (SIC) are applied [2]. NOMA is shown to have the potential of handling a massive number of connections while offering a superior sum capacity and user fairness. NOMA-based cellular networks have been projected to offer diverse data-hungry applications. The notion of NOMA in 5G cellular context was initially put forward in [3] and its superior performance was demonstrated. The attractive advantages of NOMA then spark off a substantial amount of research [4-7]. Therefore, understanding NOMA and its utilization in cellular communication systems is extremely important. In this article, we elaborate the concept of NOMA and explain how it meets some of the requirements of 5G and Beyond in order to educate the learners and researchers in the area of radio access technology.

## 2. Fundamentals of NOMA

In general, the existing NOMA schemes can be classified into two categories: power-domain NOMA and code-domain NOMA. The former assigns a unique power level to a user and multiple users transmit their signals sharing the same time-frequency-code resources, each using its allocated power [4]. The level of power for a user is decided based on its channel gain: a user with higher channel gain is often assigned a lower power level. At the receiving ends, different users' signals can be separated by exploiting the users' power-difference based on SIC. Code-domain NOMA relies on codebooks, spreading sequences, interleaving patterns, or scrambling sequences to non-orthogonally allocate resources to users [10]. Although the main focus of this article is on power-domain NOMA, basic principles of code-domain NOMA will be provided as well.

## 2.1 Power-Domain Downlink NOMA

Figure 1 presents a simple NOMA system consisting of a single BS and two users, each equipped with a single antenna. Suppose that $x_1$ and $x_2$ are the signals to be transmitted from the BS to users 1 and 2, respectively. The BS transmits the superposition coded signal as

$$s = \sqrt{P_1}x_1 + \sqrt{P_2}x_2, \tag{1}$$

where $P_i, i = 1, 2$, is the transmit power for user $i$ and the message signal $x_i, i = 1, 2$, is of unit power, i.e., $E\{|x_i|^2\} = 1$, with $E\{\cdot\}$ as the expectation operator. The total transmit power of users 1 and 2 can then be written as $P = P_1 + P_2$. In practice, for a particular system setting, $P$ is predefined and is thus divided into $P_1$ and $P_2$ according to the adopted power allocation (PA) scheme. The received signal at user $i$ can be represented as

$$y_i = h_i s + n_i, \tag{2}$$

where $h_i$ is the channel gain between the BS and user $i$ and $n_i$ represents the Gaussian noise plus interferences with power spectral density $N_{f,i}$. For a multi-cell scenario, the inter-cell interference is also included in $n_i$.

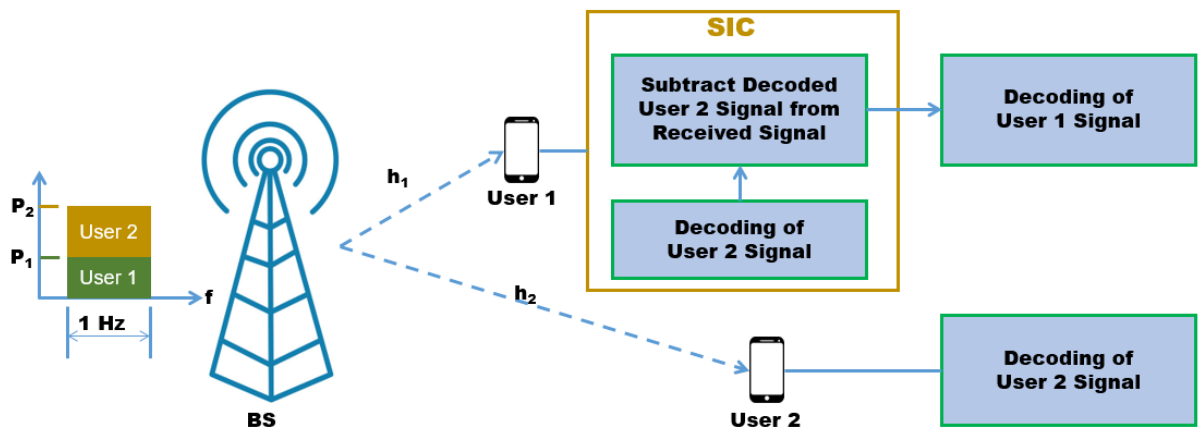

**Figure 1.** 2-user power-domain downlink NOMA.

To separate different users' signals, SIC is used at the receivers. The optimal decoding order of SIC is in the decreasing order of the strengths of the users' channels, determined by $|h_i|^2/N_{f,i}$. With this order, each user can substantially eliminate interferences from the signals of other users whose decoding orders appear after that user. Therefore, user 1 (with the maximum channel strength $|h_1|^2/N_{f,1}$), alternatively called the strong user, can cancel the interference from user 2 (with the least channel strength $|h_2|^2/N_{f,2}$), referred to as the weak user. It is worth noting that the BS periodically performs the SIC ordering based on the channel state information (CSI) feedback received from users, and the users get the updated information on the SIC ordering from the BS. Without loss of generality, it can be stated that a user with a weaker channel strength (i.e., weak user) is allocated higher power compared to a user with a stronger channel strength (strong user) to increase its signal-to-interference-plus-noise ratio (SINR). For the 2-user NOMA with $|h_1|^2/N_{f,1} > |h_2|^2/N_{f,2}$ (and hence $P_1 < P_2$), only user 1 performs SIC. It first decodes $x_2$, the signal of user 2, and subtracts it from the received signal $y_1$, after which it decodes its own signal. User 2 treats $x_1$, the signal of user 1, as noise and thus directly decodes its own signal from $y_2$ without SIC. If SIC is perfect, the achievable data rate of the NOMA user $i$, $R_i^{\text{NOMA}}$ for a transmission BW of 1 Hz can thus be written as

$$R_1^{\text{NOMA}} = \log_2\left(1 + \frac{P_1|h_1|^2}{N_{f,1}}\right), \tag{3}$$

$$R_2^{\text{NOMA}} = \log_2\left(1 + \frac{P_2|h_2|^2}{P_1|h_2|^2 + N_{f,2}}\right). \tag{4}$$

The achievable sum capacity is $R^{\text{NOMA}} = R_1^{\text{NOMA}} + R_2^{\text{NOMA}}$. Equations (3) and (4) suggest that the BS can control the data rate of each user by tuning the power allocation coefficients $\alpha_1$ and $\alpha_2$, with $\alpha_1 = \frac{P_1}{P}$ and $\alpha_2 = \frac{P_2}{P}$. To get a comparative understanding of the data rate performances of NOMA and OMA, we consider the 2-user FDMA scheme sketched in Figure 2, where the 1 Hz transmission BW is divided for the two users — user 1 uses $W$ Hz while user 2 uses the remaining $1 - W$ Hz of the BW and the power allocation ratio ($\alpha_1:\alpha_2 = P_1:P_2$) remains the same as for the NOMA scheme. Then, the achievable data rate of the OMA user $i$, $R_i^{\text{OMA}}$ can be written as

$$R_1^{\text{OMA}} = W \log_2\left(1 + \frac{P_1|h_1|^2}{WN_{f,1}}\right). \qquad (5)$$

$$R_2^{\text{OMA}} = (1-W) \log_2\left(1 + \frac{P_2|h_2|^2}{(1-W)N_{f,2}}\right). \qquad (6)$$

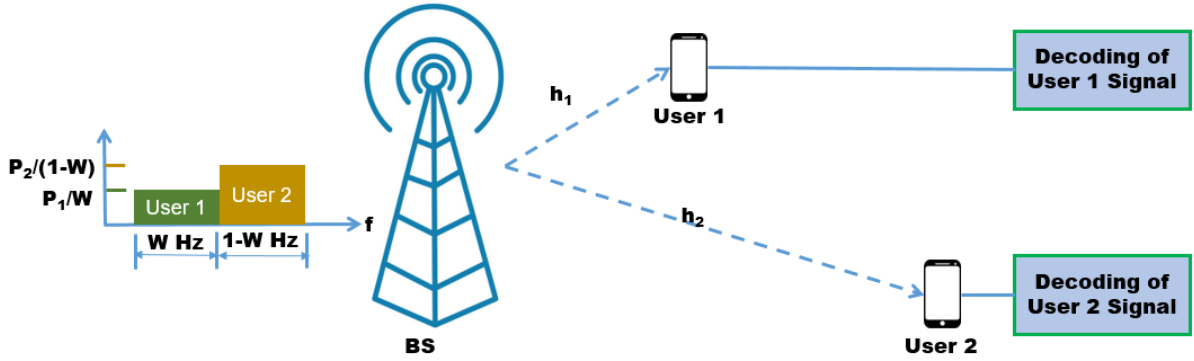

**Figure 2.** 2-user downlink FDMA.

The achievable sum capacity is $R^{\text{OMA}} = R_1^{\text{OMA}} + R_2^{\text{OMA}}$. Equations (5) and (6) suggest that no OMA user suffers from the interference from the signal of the other user, unlike NOMA as indicated by (4). In Figure 3, we present the data rate regions of both downlink NOMA and OMA by plotting the data rate of a user with respect to the data rate of the other user at different power allocation ratios with $h_1 = 10$ and $h_2 = 1$. As noticed, the rate region of NOMA is much wider compared to OMA. For example, if we expect the data rate of user 2 to be 1 bit/Hz, the achievable rate of user 1 for NOMA is approximately 9 bits/Hz, which is much higher than that for OMA with equal BW sharing. This is because the data rate of NOMA user 1 is bandwidth-limited rather than power-limited. Although a small transmission power is allocated to user 1 compared to user 2, its high channel gain (i.e., $|h_1|^2/N_{f,1}$) and the use of SIC allows the user to take advantage of the utilization of the full BW. It is possible to increase the rate of OMA user 2 if a significant fraction of BW is assigned to this user. However, it causes a severe reduction in the data rate of OMA user 1. Moreover, as presented in Figure 4, we notice that NOMA significantly outperforms OMA in terms of the sum capacity irrespective of the BW allocation ratio. Note that we considered asymmetric channel condition (different channel gains for users 1 and 2) to perform the above comparisons on the data rate regions. We can find that OMA and NOMA provide identical data rate regions in a symmetric channel environment— multiplexing in power domain does not become advantageous as the signals of both users undergo the same interference and noise [7]. Therefore, compared to OMA, NOMA in cellular downlink provides higher sum capacity and performs a better tradeoff between system efficiency and user fairness when the channel conditions are different among users. It is demonstrated that other forms of OMA techniques including TDMA and orthogonal frequency-division multiple access (OFDMA) are also significantly outperformed by NOMA in terms of throughput under different settings [8].

As a generalization of the 2-user scenario, for a $K$-user downlink NOMA with $|h_1|^2/N_{f,1} > |h_2|^2/N_{f,2} > \cdots |h_k|^2/N_{f,k} \ldots > |h_K|^2/N_{f,K}$ (and hence $P_1 < P_2 < \cdots P_k \ldots < P_K$) and 1 Hz BW, the data rate of user 1 remains the same as of (3), while the data rate of any other user $k$ is represented as

$$R_k^{\text{NOMA}} = \log_2\left(1 + \frac{P_k|h_k|^2}{\sum_{j=1}^{k-1} P_j|h_k|^2 + N_{f,k}}\right). \tag{7}$$

Although theoretically there is no limitation on the number of NOMA users, from a practical viewpoint, NOMA in downlink is applied to a small number of users (typically two or three) in a cluster; for a large number of users, there is a degradation in the bit error rate due to error propagation primarily originated from imperfect SIC. Also, as the number of users in a cluster increases, end user devices require more computing power and higher energy, which might not be feasible in practice, especially for resource-constrained devices. Also, it is worth noting that power-domain NOMA in downlink has been standardized in the 3rd Generation Partnership Project (3GPP) LTE Release 13, referred to as multiuser superposition transmission (MUST) [45].

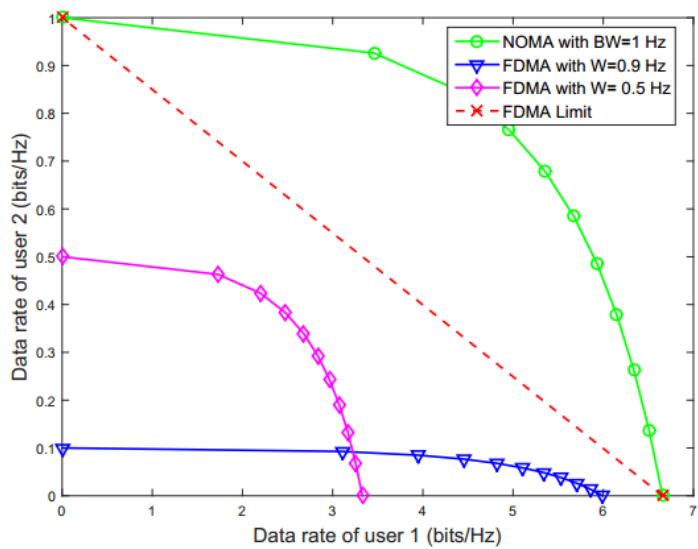

**Figure 3.** Rate region in 2-user downlink NOMA and OMA.

## 2.2 Power-Domain Uplink NOMA

Figure 5 presents an uplink NOMA scheme, where users 1 and 2 simultaneously transmit their signals $x_1$ and $x_2$ to the BS. The received signal at the BS is given by

$$y = \sum_{i=1}^{2} \sqrt{P_i}\, h_i\, x_i + n, \tag{8}$$

where $P_i$ is the transmit power for user $i$, with $E\{|x_i|^2\} = 1$, and $n$ represents the Gaussian noise plus interferences with power spectral density $N_f$. In general, the BS broadcasts a downlink reference signal based on which each user performs channel estimation. Thus, they can regulate their transmit power to either $P_1$ or $P_2$, depending upon their channel gains. User 1 is again marked as the strong user experiencing a higher channel gain compared to user 2, the weak user. Upon receiving the superposed signal, the BS performs SIC to separate the users' signals. The BS first decodes the signal of user 1 by treating the signal of user 2 as noise [16], then subtracts the decoded signal of user 1 ($\hat{x}_1$) from the received signal ($y$). From the remaining, the signal of user 2 ($\hat{x}_2$) is then decoded. Therefore, in uplink NOMA, user 1 experiences interference from user 2, whereas user 2 receives no interference from user 1 since user 1's signal is removed prior to decoding user 2's signal. In contrast, user 2 in downlink NOMA experiences interference from user 1, whereas user 1 does not suffer interferences from user 1 since user 2's signal is removed prior to decoding user 1's signal. If SIC is perfect, the achievable data rate of the NOMA user $i$, $R_i^{\text{NOMA}}$ for a transmission BW of 1 Hz is expressed as

$$R_1^{\text{NOMA}} = \log_2\left(1 + \frac{P_1|h_1|^2}{P_2|h_2|^2 + N_f}\right), \tag{9}$$

$$R_2^{\text{NOMA}} = \log_2\left(1 + \frac{P_2|h_2|^2}{N_f}\right). \tag{10}$$

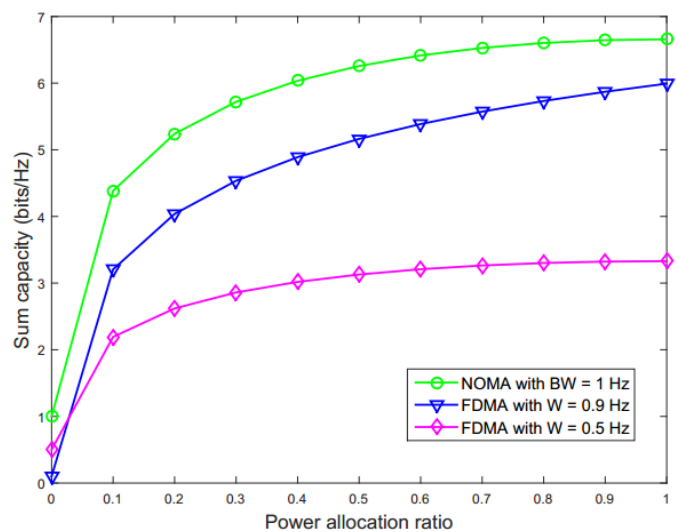

**Figure 4.** Sum capacity in 2-user downlink NOMA and OMA.

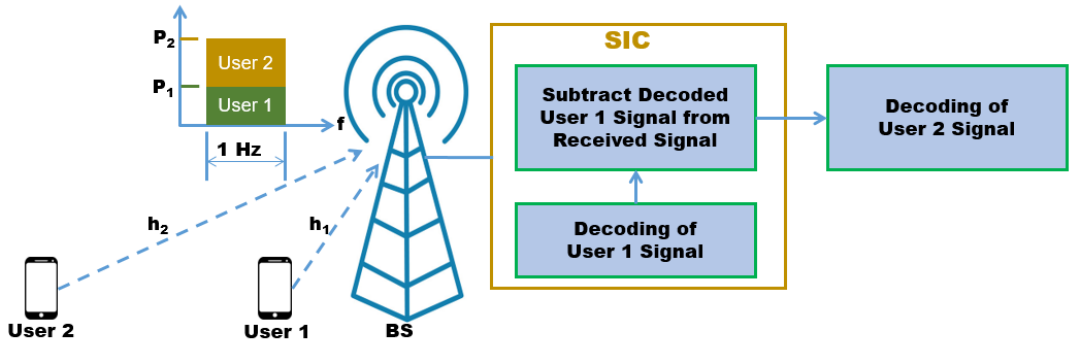

**Figure 5.** 2-user power domain uplink NOMA.

The achievable sum capacity is $R^{\text{NOMA}} = R_1^{\text{NOMA}} + R_2^{\text{NOMA}}$. One can easily understand that the same equations (5) and (6) can also be utilized to calculate the achievable data rates of the uplink OMA users 1 and 2 after replacing $N_{f,i}$ with $N_f$. Like downlink NOMA, uplink NOMA can also be shown to be superior in terms of achievable data rate and sum capacity by comparing (9) and (10) with their OMA counterparts [7].

As a generalization of the 2-user case, for a $K$-user uplink NOMA with $|h_1|^2/N_f > |h_2|^2/N_f > \cdots |h_k|^2/N_f \ldots > |h_K|^2/N_f$ (and hence $P_1 < P_2 < \cdots P_k \ldots < P_K$) and 1 Hz BW, the data rates of user $k$, $k = 1, 2, \ldots, K-1$, and $K$ can be written as

$$R_k^{\text{NOMA}} = \log_2\left(1 + \frac{P_k|h_k|^2}{\sum_{j=k+1}^{K} P_j|h_j|^2 + N_f}\right), \tag{11}$$

$$R_K^{\text{NOMA}} = \log_2\left(1 + \frac{P_K|h_K|^2}{N_f}\right). \tag{12}$$

Unlike downlink NOMA, uplink NOMA can accommodate a relatively larger number of users because BS can usually be equipped with required computing power and energy. Moreover, BS can apply more computing-intensive decoding schemes to reduce the impact of interferences. Therefore, uplink NOMA is preferable to downlink NOMA for massive machine type communications [9].

**2.3 Code-Domain NOMA**

In contrast to power-domain NOMA, the code-domain NOMA achieves multiplexing in the code domain. The notion of the latter is motivated by the classic CDMA systems, in which several users share the same time-frequency resources, while utilizing unique user-specific spreading sequences. However, compared to CDMA, the unique feature of the code-domain NOMA is that the spreading sequences are limited to sparse sequences (alternatively called low-density sequences) or non-orthogonal low cross-correlation sequences.

The most basic form of code-domain NOMA is the low-density spreading-CDMA (LDS-CDMA) [11]. To understand how LDS-CDMA works, we start with a brief discussion on the working principle of a classical CDMA system. In a CDMA system, a user-specific high-rate pseudo-noise spreading sequence is first generated. The sequence is then divided into a number of chips. Next, the low-rate symbol of the user is spread over each chip before transmission. At the same time, other users in the system also transmit their symbols by using their respective spreading sequences. At the receiving side, the signals received during all the chips need to be combined to separate the different signals by despreading the signals using the same spreading sequences as transmission. Here, the elements of the spreading sequences are generally non-zero, i.e., the spreading sequences are not sparse. Therefore, the signals received from all the active users are overlaid on top of each other at each chip, and each user will experience inter-user interference imposed by all the other users. The CDMA system described above is called non-orthogonal type, where the spreading sequences are non-orthogonal. On the other hand, orthogonal CDMA systems use orthogonal spreading sequences and it is straightforward to remove the interferences. Therefore, a simple correlation receiver is sufficient to detect the signal of a user. However, an orthogonal CDMA system can support only as many users as the number of chips. In contrast, the non-orthogonal CDMA system has many more codes than the number of chips in a sequence, but since the codes are non-orthogonal, they suffer from inter-user interferences. Hence, the receiving side in an non-orthogonal CDMA system requires more complex multi-user detection (MUD)s. On that, the idea of LDS-CDMA emerges, which uses sparse spreading sequences (i.e., the number of non-zero elements in the spreading sequence is much lower than the number of chips) instead of the classic "fully-populated" spreading sequences to support more users than orthogonal counterpart with reduced interference. Since all symbols are modulated onto sparse spreading sequences before transmission, each user will only spread its data over a small number of chips resulting in a less number of superposed signals (and hence, reduced amount of effective interferences) at each chip compared to the number of active users. Therefore, the use of low-density spreading sequences enables LDS-CDMA to offer improved system performances by using low-density spreading sequences, when compared to conventional CDMA.

Another family of code-domain NOMA is LDS-OFDM, which can be thought of as an amalgamation of LDS-CDMA and orthogonal frequency-division multiplexing (OFDM) [12]; the original data streams are first multiplied with the low-density spreading sequences and then transmitted over different subcarriers. The operation of LDS-CDMA thus becomes analogous to multi-carrier (MC) CDMA (MC-CDMA), keeping all the benefits of OFDM-based MC transmissions in terms of its inter symbol interference avoidance, together with MUD-mediated LDS-CDMA operating at a lower complexity. Sparse code multiple access (SCMA) [13] is a recent code-domain NOMA technique based on LDS-CDMA. In contrast to LDS-CDMA, the procedure of bit to quadrature amplitude modulation symbol mapping and spreading are combined together and input data streams are directly mapped into a multidimensional codeword of an SCMA codebook set. Compared to LDS-CDMA, SCMA provides a low-complexity reception technique and offers better performances. The reader is referred to [10] for a comprehensive discussion of the code-domain NOMA techniques, which highlights both transmitter schemes and multi-user detectors at the receiver, emphasizing their advantages and drawbacks.

## 2.4 Advantages of NOMA

Compared to OMA, some of the key advantages offered by NOMA are summarized as follows:

*High Spectral Efficiency:* Since it can serve multiple users by employing the same resource block, NOMA is highly spectrum-efficient, and hence, improves the system throughput.

*Massive Connectivity:* Massive IoT is in reference to massive scale, billions of devices, objects and machines that need connectivity even in the most remote locations. The 5G and Beyond are expected to support this massive connectivity and NOMA can fulfill the expectation in some scale. Since the number of supportable NOMA users/devices is not strictly limited by the number of available orthogonal resources, NOMA is capable of serving them by using less resources.

*Improved User Fairness:* The power allocation of NOMA allows a system to perform a tradeoff between fairness among users and throughput [14]. Therefore, if an appropriate power allocation is adopted, the cell-edge users can also enjoy higher data rates while maintaining the system throughput.

*Low latency:* OMA depends on access-grant requests — an uplink user first has to send a scheduling request to the BS. The BS then sends a clear signal to the user in the downlink channel. The access-grant process thus increases latency (even with additional signaling overhead), which is not desirable on the next generation connectivity. Additionally, with massive connectivity, orthogonal pilots do not suffice and the access-grant procedure becomes even more complex. In contrast, grant-free multiple access can be realized in uplink NOMA schemes by blindly detecting active users and decoding their data streams [15]. The grant-free NOMA thus comes with reduced latency. Further, it should be mentioned that a reduction in latency can be obtained, as multiple users are simultaneously served in NOMA.

## 3. Multiple-input Multiple-output NOMA (MIMO-NOMA)

The application of MIMO to NOMA is of significance, since MIMO introduces extra spatial degrees of freedom for system performance improvement. Research on MIMO-NOMA has attracted great attention from both academia [46-50] and industry [7, 51].

### 3.1 Downlink MIMO-NOMA

Compared with single-input single-output NOMA (SISO-NOMA), the main challenge in MIMO-NOMA comes from the fact that the MIMO channel is non-degraded, i.e., users cannot be ordered based on their channel strengths in general settings. As a result, MIMO-NOMA is generally not capacity-achieving. Indeed, the complex dirty paper coding (DPC) is the only strategy that achieves the capacity region of the multiple-input single-output (MISO) (Gaussian) broadcast channel with perfect CSI at the transmitter. Nonetheless, under certain conditions, MIMO-NOMA can still achieve the same performance as DPC. In [46], the concept of quasi-degradation is introduced for a two user MISO-NOMA system with minimum rate constraint, and it is shown that NOMA can achieve the same performance as DPC if the channels are quasi-degraded.

Meanwhile, user ordering is a difficult task for MIMO-NOMA. Unlike SISO-NOMA, in which user channels are scalars and can be easily ordered, in MIMO-NOMA, user channels are in the form of vectors or matrices and cannot be ordered directly. A simple way to handle this is to order the users just using the large-scale path loss. However, this may yield system performance degradation, since small-scale channel information is not exploited. To fully recap the spatial degrees of freedom, conceiving an appropriate beamforming/precoding design is essential for MIMO-NOMA systems. In particular, both the power domain and the angular domain should be considered in beamforming design to enhance the system spectral efficiency. There exist two popular MIMO-NOMA designs: 1) cluster-based MIMO-NOMA design [47-50, 52-53]; and 2) beamforming-based MIMO-NOMA design [46, 54-56].

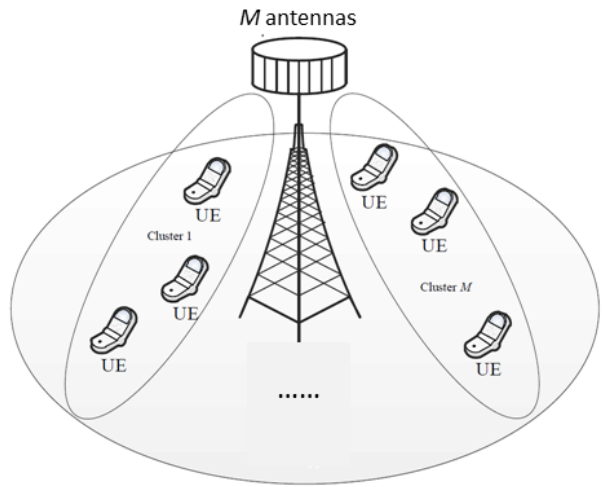

**Figure 6.** Illustration of cluster-based MIMO-NOMA.

**Cluster-based MIMO-NOMA:** As shown in Figure 6, the cluster-based MIMO-NOMA design works as follows: users are first grouped into clusters based on their channels; then, the users allocated to the same cluster share a common beamforming vector. Note that deriving the optimal user clustering often requires to exhaustively enumerate all possible user clustering combinations and is computationally prohibited. A good heuristic for this is to put users with high channel correlation and gain difference into the same cluster, since high channel correlation facilitates the beamforming design while high gain difference favors the implementation of SIC [52]. Once users are clustered, the main task lies in how to design the appropriate beamforming vectors for each cluster. The simplest beamforming design is to use random beamforming, which is also effective in reducing the CSI feedback [47, 53]. In [53], the authors propose to combine random beamforming with intra-beam SIC for downlink NOMA transmission. However, due to the existence of inter-cluster interference, different clusters are still coupled. To address this issue, the authors in [47] propose to use random beamforming at the BS, and zero-forcing detection at the user side. Then, inter-cluster interference can be removed at each user, and the MIMO-NOMA system is decomposed into a set of independent SISO-NOMA arrangements. To illustrate how this works, let us consider a system with $2M$ users, which are randomly grouped into $M$ pairs. Denote the two users in pair $i$ by user $i_1$ and $i_2$, respectively. Meanwhile, it is assumed that the BS is equipped with $M$ antennas, while the users are equipped with $N$ antennas. The received signal at user $i_k$, $k = 1,2$, is given by

$$\mathbf{y}_{i_k} = \mathbf{H}_{i_k}\mathbf{Px} + \mathbf{n}_{i_k}, \tag{13}$$

where $\mathbf{H}_{i_k} \in C^{N \times M}$ is the channel matrix, while $\mathbf{P} = \mathbf{I}_M$ denotes the precoding matrix. Besides, $\mathbf{x} = [x_1, \dots, x_M]^T$, where $x_i$ is a NOMA mixture containing the two symbols for the two users in the same pair. Denote the zero-forcing detection vector for user $i_k$ by $\mathbf{v}_{i_k}$, satisfying $\mathbf{v}_{i_k}^H \mathbf{h}_{m,i_k} = 0$, for $m \neq i$, where $\mathbf{h}_{m,i_k}$ is the $m$th column of $\mathbf{H}_{i_k}$. By applying $\mathbf{v}_{i_k}$ to the received signal $\mathbf{y}_{i_k}$, we can obtain

$$\mathbf{v}_{i_k}^H \mathbf{y}_{i_k} = \mathbf{v}_{i_k}^H \mathbf{h}_{i,i_k} + \mathbf{v}_{i_k}^H \mathbf{n}_{i_k}. \tag{14}$$

Now it can be seen that each cluster can be handled separately. Without loss of generality, we assume that $|\mathbf{v}_{i_1}^H \mathbf{h}_{i,i_1}| \geq |\mathbf{v}_{i_2}^H \mathbf{h}_{i,i_2}|$. On this basis, we have

$$\begin{aligned} R_{i_1}^{i_2} &= \log\left(1 + \frac{\rho \alpha_{i_2} |\mathbf{v}_{i_1}^H \mathbf{h}_{i,i_1}|^2}{\rho \alpha_{i_1} |\mathbf{v}_{i_1}^H \mathbf{h}_{i,i_1}|^2 + 1}\right) \\ &\geq \log\left(1 + \frac{\rho \alpha_{i_2} |\mathbf{v}_{i_2}^H \mathbf{h}_{i,i_2}|^2}{\rho \alpha_{i_1} |\mathbf{v}_{i_2}^H \mathbf{h}_{i,i_2}|^2 + 1}\right) = R_{i_2}, \end{aligned} \tag{15}$$

where $R_{i_1}^{i_2}$ and $R_{i_2}$ denote the achievable rate of user $i_2$'s signal at users $i_1$ and $i_2$, respectively. $\rho$ is the transmission signal-to-noise ratio, while $\alpha_{i_1}$ and $\alpha_{i_2}$ are the power coefficients for users $i_1$ and $i_2$, respectively. This shows that the interference from the weak user $i_2$ can always be removed at the strong user $i_1$, just as in SISO-NOMA. Note that only two users for each cluster are considered here. Interested readers can refer to [49] for the general scenario with multiple users in each cluster. Moreover, in [49] it is further proved that the capacity achieved by MIMO-NOMA always exceeds that obtained by MIMO-OMA. Nonetheless, [47] and [49] only apply to the scenario when the number of receive antennas exceeds that of transmit antennas, i.e., $N \geq M$. To relieve this constraint, the so-called signal alignment technique is introduced in [48]. The basic idea of signal alignment is to set the detection vectors $\mathbf{v}_{i_1}$ and $\mathbf{v}_{i_2}$ such that $\mathbf{v}_{i_1}^H \mathbf{h}_{i_1} = \mathbf{v}_{i_2}^H \mathbf{h}_{i_2}$. On this basis, the two users in the same cluster can be treated as a single user, and then, conventional zero-forcing precoder can be used as in MIMO-OMA, i.e., $\mathbf{P}_i$ should satisfy

$$[\mathbf{v}_{1_k}^H \mathbf{h}_{1_k}, \dots, \mathbf{v}_{i_k}^H \mathbf{h}_{i_k}, \dots \mathbf{v}_{M_k}^H \mathbf{h}_{M_k}]_{2(M-1) \times M} \mathbf{P}_i = 0. \tag{16}$$

With such choices of $\boldsymbol{P}_i$ and $\boldsymbol{v}_{i_k}$, a SISO-NOMA model similar to that in (14) is obtained. Moreover, the constraint $N \geq M$ required in [47] can be relaxed to $N > M/2$. In [50], it is further shown that MIMO-NOMA always outperforms MIMO-OMA in terms of sum rate. Figure 7 shows the ergodic sum rate comparison between MIMO-NOMA and MIMO-OMA. It is clear that NOMA outperforms OMA, and the gap increases with the channel gain difference.

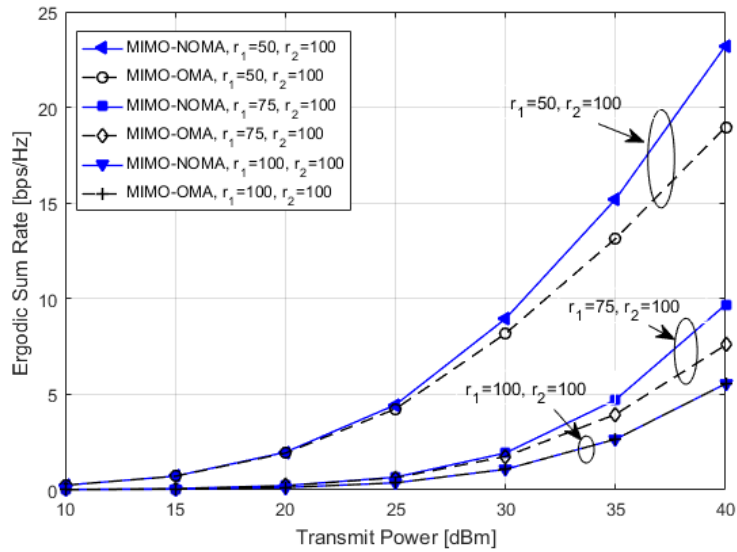

**Figure 7.** Ergodic sum rate vs. transmit power, for different distance values [50, Figure 2].

**Beamforming-based MIMO-NOMA:** As shown in Figure 8, users are no longer grouped into clusters in this case. Instead, each user is assigned its own beamforming vector. However, unlike MIMO-OMA, SIC is still performed at the user side to remove the interference. To guarantee successful SIC, the following constraint should be explicitly given, i.e., the achievable rate at each user cannot exceed the minimum rate among all users which need to decode it. Note that as shown in [49], this constraint is less of an issue in clustered-based MIMO-NOMA, since users in the same cluster share a common beamforming vector. To show this, we consider a scenario with $K$ users, and assume that the ascending order is adopted among the users for decoding (note that other user ordering can also be employed here). Then, for user $k$, its signal should be decoded and removed by users $k$+1 to $K$. Denote the achievable SINR threshold for user $k$ as $\gamma_k$, the constraint can be expressed as follows:

$$\gamma_k \leq min(SINR_k^k, \ldots, SINR_K^k), k \in \{1, \ldots, K\}, \tag{17}$$

where $SINR_n^k, n = k, \ldots, K$ denotes the SINR of user $k$'s signal on user $n$.

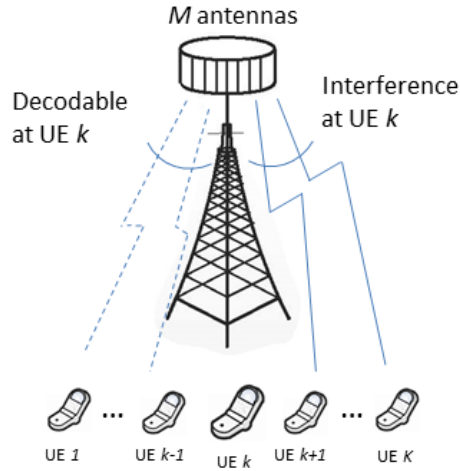

**Figure 8:** Illustration of beamforming-based MIMO-NOMA.

In [54], the ergodic sum rate maximization problem is studied for a Rayleigh fading based MIMO-NOMA systems with statistical CSI at the transmitter. Both optimal and low-complexity suboptimal power allocation schemes are proposed under total transmit power constraint and minimum rate constraint of the weak user. Numerical results show that the proposed NOMA schemes significantly outperform traditional OMA. In [55], a layered transmission scheme is proposed based on QR factorization. Under instantaneous CSI, an approach to maximize the sum rate of MIMO-NOMA with layered transmissions is proposed after showing that the sum rate is concave in allocated powers to multiple layers of users. Under statistical CSI, a closed-form expression for the average sum rate is derived, and on this basis, power is allocated using alternating optimization. In [46], optimal precoding is studied for a quality-of-service (QoS) constrained optimization problem in a MISO-NOMA network. It is shown that if the broadcast channels are quasi-degraded, the proposed optimization algorithm in combination with superposition coding and SIC can achieve system capacity. Note that [46, 54-55] only apply to two users. In [56], the general scenario with multiple users is considered. The sum rate maximization is handled with the help of minorization-maximization algorithm.

To summarize, cluster-based MIMO-NOMA is easier to handle, while beamforming-based MIMO-NOMA may achieve better system performance.

### 3.2. Uplink MIMO-NOMA

In an uplink system, where a multiple antenna BS serves $K$ single antenna users, denote the antenna number at the BS by $M$. The received signal at the BS is given by

$$\boldsymbol{y} = \sum_{k=1}^{K} \sqrt{P_k} \boldsymbol{h}_k x_k + \boldsymbol{n}, \tag{18}$$

where $\boldsymbol{h}_k \in C^M$ is the channel vector between user $k$ and BS. Moreover, $x_k$ and $P_k$ denote the signal and transmit power for the $kth$ user, respectively, while $\boldsymbol{n}$ represents the additive white Gaussian noise vector.

At the BS, the receiver of a minimum mean square error (MMSE)-based linear filtering followed by a SIC (MMSE-SIC) can be employed for signal detection [7]. On this basis, it is shown in [7] that the sum rate of the $K$ users is independent from the user ordering, and equals the system capacity. Nonetheless, the individual rate is affected by the user ordering, and users decoded at a later stage of the SIC process can benefit from increased throughput. To enhance system fairness, users with better channel conditions are decoded earlier.

Although the MMSE-SIC receiver can achieve the system capacity in theory, it suffers from error propagation in practical implementation. To relieve this, other detection methods can also be used, e.g., zero-forcing. In [57], zero-forcing is employed among the strong users for a millimeter wave (mmWave) massive MIMO system. Moreover, an enhanced scheme is proposed to further remove inter-cluster interference using SIC. Note that this enhanced scheme can be considered as a middle way between MMSE-SIC and zero-forcing.

### 4. Resource Allocation

The main resources in wireless communication systems include time, frequency, space, code and power. In NOMA systems, multiple users are accommodated in each time/frequency/code resource block (RB), forming a NOMA cluster. As a result, how to group the users into NOMA clusters and allocate the power is of significance. In the following, we will discuss user clustering and power allocation in various NOMA systems in detail.

### 4.1 User clustering

In general, user clustering is a hard problem, owing to the inherent combinatorial nature [58]. Indeed, for downlink, it has been shown that assigning NOMA users to different orthogonal RBs, e.g., subcarries/subchannels, belongs to a non-deterministic polynomial-time (NP)-hard problem [58-60]. To make it tractable, a single two-user pair is thoroughly investigated in [61], showing that the performance gain of NOMA with fixed PA (F-PA) over OMA grows with increasing the difference between the channel gains of the users of interest. This positive correlated relation between performance gain and channel gain difference has been widely exploited for proposing greedy user clustering methods. Specifically, in [61], the authors propose to pair the user of highest channel gain with the user of lowest channel gain, the user of second highest channel gain with the user of second lowest channel gain and so on. However, this may lead to unfair distribution of the performance gain over different clusters. To address this issue, [62] proposes to first divide the users into two groups based on their channel gains. Then, the user of highest channel gain in group one is paired with its counterpart in group two, and so on. Moreover, it has been proved that this user clustering method can yield the optimal solution when combined with appropriate power allocation [63]. Nonetheless, [63] can only be applied to frequency flat channels.

These channel gain-based user clustering methods have low-complexity. However, their performance may be unstable, since they are heuristic methods. To strike a balance between complexity and effectiveness, systematic frameworks should be adopted. In the following, we list some promising approaches:

**Monotonic optimization:** The resource allocation problem in NOMA systems is generally non-convex due to the existence of intra-cluster interference. As a result, obtaining the optimal solution using the convex optimization theory is rather challenging. Therefore, other properties except for convexity should be exploited. Among them, the monotonicity is an important property, which can be used to address non-convex problems. In [64], the authors use monotonic optimization to develop an optimal solution for the joint power and subcarrier allocation problem.

**Combinatorial relaxation:** The main challenge of user clustering comes from the binary variable of associating a user to its corresponding cluster. By relaxing this binary variable into continuous one, the original NP-hard problem often can be transferred into a convex problem, for which the optimal solution can be obtained using convex optimization. Note that the binary variables should be recovered, using techniques like rounding. However, this kind of relaxation and recovering often yield a performance gap between the original problem and the relaxed one.

**Game theory and matching theory:** Recently, the game theory has been widely used for user clustering in NOMA systems [65-67]. By treating users as players, and RBs as action profile, a coalition game is formed in [65]. Then, a user clustering is proposed based on the preference relation. Note that the traditional coalition game is adopted in [65], i.e., the strategy of each user is to maximize its own utility rather than improving the system performance. To overcome this, an improved coalition game is proposed in [66] by following the particle swarm optimization method which adjusts the utility function for each user towards a global optimal solution. Except for the coalition game, a Stackelberg game approach is proposed in [67] for a mmWave NOMA system, by considering user clustering as the leader, and power allocation as the follower. The shortcomings of game theory-based approaches lie in the distributed implementation and the unilateral equilibrium deviation.

Matching theory, as a powerful mathematical tool for handling the combinatorial user allocation problems, is capable of overcoming these issues [68]. However, due to the existence of intra-cluster interference, conventional matching algorithms, e.g., the Hungarian algorithm, cannot be used for NOMA systems [69]. Instead, many-to-many two-sided matching theory is invoked for resource allocation in NOMA systems, and then, iterative swap-based matching algorithms are proposed [60, 69-70].

The computational complexity of the above approaches may be too high for practical implementation. To address this issue, a good practice is to exclude the user groups that are not appropriate for NOMA multiplexing from unnecessary comparison of candidate user pairs [3]. In addition, the previous user clustering algorithms do not consider the fact that there might not be enough strong users for pairing. In this case, there may exist leftover weak users after each strong user is paired with its partner. To handle this, a hybrid approach can be adopted, where the leftover users are accessed via OMA. However, this deprives these users of the gains provided by NOMA. Alternatively, NOMA can still be implemented employing the concept of virtual user clustering [71], where a frequency band is shared by two weak users and a strong user: half of the bandwidth is used by the strong user and a weak user, while the other half is used by the strong user and the other weak user.

**4.2. Power Allocation in NOMA**

PA plays a pivotal role in NOMA, as users are multiplexed in the power domain [9]. It directly impacts the system performance, such as interference management, rate distribution, and even user admission. An inappropriate PA could lead to not only an unfair rate distribution among users, but also system outage due to SIC failure. When designing PA strategies, one needs to consider users' channel conditions, availability of CSI, QoS requirements, total power constraint, system objective and so on. Some widely adopted PA performance metrics include the number of admitted users, sum rate, energy efficiency, user fairness, outage probability and total power consumption. Therefore, the goal of PA in NOMA is to achieve either more admitted users, higher sum rate and energy efficiency (EE), or a balanced fairness under minimum power consumption. A variety of PA strategies have been proposed in the literature, targeting various aspects of PA in NOMA, and a classification is provided in Figure 9. We introduce PA in the following two subsections: one focuses on single-carrier (SC) systems, while the other deals with MC systems.

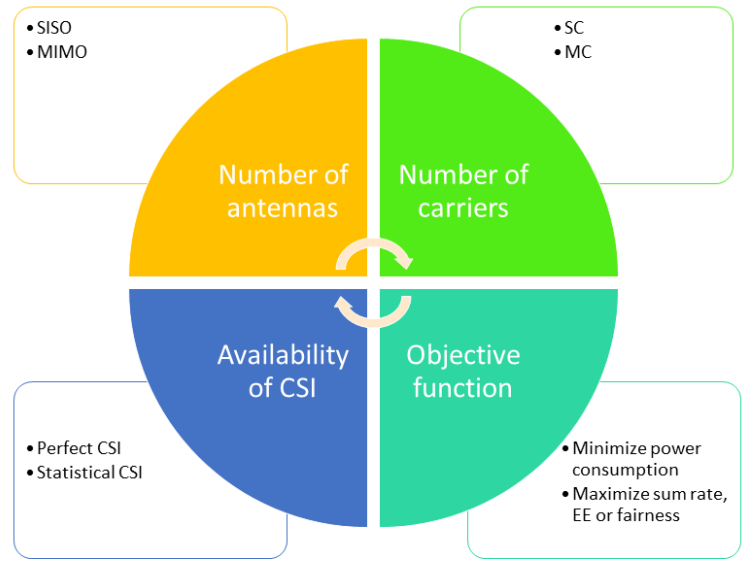

**Figure 9:** Classification of PA strategies [62, Figure 4].

**PA in SC-NOMA:** In downlink SC-NOMA systems, the optimal PA to maximize the sum rate simply allocates all power to the user with the best channel [59, 72-73]. Clearly, this results in extreme unfairness among users and decreases the number of admitted users as well. To strike a balance between system throughput and user fairness, more power is allocated to the weak user in NOMA. By doing this, the strong user can remove the interference from the weak user via SIC, while its interference to the weak user remains comparatively small. F-PA is the simplest PA algorithm, and it allocates power to the users ultilizing a fixed ratio based on their positions in the channel ordering. Since users' specific channel gains are not exploited during PA, F-PA may not meet users' various QoS requirements. To handle this, fractional transmit power control allocates power to the users inversely proportional to their channel gains powered with a decaying factor. Nonetheless, assigning the same decaying factor to all users is still suboptimal, and how to select the appropriate decaying factor to balance system throughput and user fairness is an open issue.

The availability of CSI directly impacts the PA in NOMA. Under perfect CSI, the authors in [59] show that the weighted sum rate maximization problem is convex, and obtain optimal PA via convex optimization. The max-min fairness problem is proved to be quasi-convex, and thus, optimal PA can be attained using the bisection method [14, 74]. The energy-efficient PA problem can be formulated as a fractional problem, for which the Dinkelbach's algorithm can be applied [37, 75-76]. Under statistical CSI, the min-max outage probability problem under a given SIC order is non-convex. Nonetheless, optimal PA is derived in [14], and based on this, the authors in [77] derive the optimal SIC decoding order.

The works above may not guarantee a higher throughput of NOMA over OMA. To ensure this for the weak user, cognitive radio (CR)-inspired PA can be deployed, in which the weak user is viewed as a primary user in a CR network [17, 47]. However, this is achieved at the expense of the strong user as it is served only after the weak user's QoS is satisfied. To address this issue, dynamic PA is proposed in [78], which allocates power to the users such that NOMA achieves strictly higher individual user rate than OMA.

**PA in MC-NOMA:** For multi-user systems, NOMA is usually integrated with MC (MC-NOMA) to reduce decoding complexity [79]. In MC-NOMA, a user can occupy multiple sub-carriers, and vice versa. MC-NOMA is quite suitable for 5G and Beyond since it is difficult to find continuous wide bandwidth in 5G and Beyond. When compared with OFDMA, MC-NOMA not only increases the system spectral efficiency, but also supports a larger number of users. Its performance is affected by both PA and sub-carrier assignment (SA). For downlink, the weighted sum rate maximization problem is proved to be NP-hard [58-60]. In contrast, for uplink, the sum rate problem is shown to be convex [80], and further, an optimal and low-complexity iterative water-filling algorithm is proposed. This major difference between downlink and uplink is because the BS decodes all user signals in uplink while each user decodes its own signals separately in downlink.

It is worth mentioning that perfect CSI is assumed in the above schemes, which might be impractical for MC-NOMA systems overloaded with exceedingly number of users. To overcome this, the resource allocation under statistical CSI should be studied. Without perfect CSI, an explicit SIC decoding order should be derived first, as the BS cannot decide the SIC decoding order directly [81]. On this basis, PA and SA can be conducted as for the case with perfect CSI. To further increase the spectral efficiency of MC-NOMA systems, full duplex (FD) BS can be deployed [82], yielding a substantial throughput enhancement when compared with FD MC-OMA and half duplex (HD) MC-NOMA systems [42].

## 5. Cooperative NOMA (C-NOMA)

Because of its attractive benefits, as mentioned in the early part of this article, substantial research has been conducted in order to adopt NOMA techniques for various systems and applications. In line with that, cooperative NOMA (C-NOMA) is one of the most active area of investigation. The concept of a C-NOMA scheme in the form of a downlink system was first presented in [17] by exploiting the prior information available in a NOMA system. Here, we first present the basic form of a C-NOMA. Then, some variants of C-NOMA are also discussed.

**Basic C-NOMA:** Consider a downlink NOMA system consisting of one BS and $K$ users. The C-NOMA introduced in [17] works in two phases, as follows. In this first phase, called direct transmission phase, the BS broadcasts the superposed message to the $K$ users by superposing $K$ messages based on the NOMA principle. Then, the messages will be decoded by the users. Note that SIC will be performed by each user except user $K$ (the weakest user). In the second phase, called cooperative phase, the users cooperate with each other via short range communication techniques such as Bluetooth and ultra-wideband. This phase consists of $(K-1)$ time slots. During the first time slot, the strongest user (i.e., user 1) first broadcasts the superposed messages to the other $(K-1)$ weak users by superposing the $(K-1)$ messages. Then, user 2 successively decodes the messages of the weak users (user 3 to user $K$) by combining the observations from the first phase and the first time slot of the second phase via diversity combining (e.g., maximum ratio combining) and eventually decodes its own message using the NOMA principle. Similarly, during the second time slot, the second strongest user (i.e., user 2) broadcasts the superposed messages to the other $(K-2)$ weak users by superposing the $(K-2)$ messages and user 3 successively decodes the messages of the weak users (user 4 to user $K$) by combining the observations from the first phase and the first and second time slots of the second phase via diversity combining and eventually decodes its own message using the NOMA principle. The process is repeated at remaining time slots. One can note that the power allocation coefficients during each phase and even at each time slot would be different based on the local channel conditions. It is intuitively understandable that the SINR of the users will be larger because of combining the signals from both phases. The use of the cooperative phase in C-NOMA can boost reception reliability. Therefore, C-NOMA outperforms non-cooperative NOMA (i.e., NOMA without the cooperative phase) in terms of both outage probability and outage capacity, since it achieves the diversity gain for all users. Considering all users in the network to take part in cooperative NOMA is however not viable because the user cooperation comes with huge computational complexity and the additional time slots would increase communication latency. Thus, user grouping can be performed to reduce the impact of the said shortcomings. Accordingly, the users in a cell can be divided into multiple groups. Then, C-NOMA can be applied within each group separately.

**Cooperative Relaying Scheme-based NOMA (CRS-NOMA):** The C-NOMA concept discussed above can be viewed as relay-assisted NOMA, since the strongest user at each time slot in the cooperative phase acts as a relay for other users. In fact, cooperative relaying systems that exploit spatial diversity represents a popular communication paradigm [18]. With a half-duplex relay, the destination in the cooperative relaying system receives and combines two independent copies, one directly from the source and the other one from the relay, of the same transmitted signal. The overall SINR at the receiver thus becomes high and improves the reception quality of the signal. However, cooperative relaying suffers from a spectral efficiency loss because of the duplication in transmissions. Therefore, a cooperative relaying scheme (CRS) using NOMA, called CRS-NOMA, is proposed to improve the overall spectral efficiency [19]. In the CRS-NOMA, the source broadcasts the superposition coded signal in the first time slot. Then, the relay performs SIC and decodes its own symbol and the symbol to be relayed for destination. In the subsequent time slot, the relay retransmits the decoded signal to the destination with full transmit power. The CRS-NOMA outperforms the conventional cooperative relaying [18] in the high SNR region in terms of sum capacity. Whereas the CRS-NOMA introduced in [19] considers Rayleigh fading channels, its performance over Rician fading channels has been investigated in [20]. A variation of the CRS-NOMA scheme is proposed in [21], in which the destination applies diversity combining and another SIC to jointly decode its signal transmitted by the source. This CRS-NOMA with joint detection outperforms the basic CRS-NOMA [19] when the channel strength between the source and relay is stronger than that of the relay to destination. C-NOMA schemes explored in [17, 19-21] have been analyzed in downlink settings. The advantage of C-NOMA has also been shown in uplink scenario [22].

**Uplink and Downlink C-NOMA:** An interesting cooperative relaying using NOMA has been investigated in [23], in which multiple sources simultaneously use a common relay to transmit their messages to the respective destinations. In this scheme, the relay first receives the transmitted symbols from the sources in an uplink NOMA manner, and then broadcasts the superposition coded signal to the destinations following the downlink NOMA principle. On that, [23] exploits the concepts of both uplink and downlink NOMA to achieve a higher ergodic capacity under cooperative scenario. The spectral efficiency of the joint uplink and downlink NOMA can be further improved by transmitting a new symbol from the source to the destination in the cooperative phase [24]. As such, in the first phase, a source broadcasts the superposing coded signal to a destination and relay in downlink NOMA mode. In the subsequent phase, whereas the relay forwards the SIC-decoded symbol to the destination, the source simultaneously transmits a new symbol for the destination following the uplink NOMA principle.

To date, there exist various forms of C-NOMA. For example, the use of a dedicated relay can enhance the transmission reliability for NOMA users with poor channel conditions [25]. Even, a dedicated relay can be useful for reliable communications to serve NOMA users equipped with multiple antennas [26]. In case of using multiple relays in a cooperative network, relay selection plays a pivotal role for robust communications. In conventional cooperative communication, the max-min criterion [27] is one of the most popular strategies for relay selection. In C-NOMA, the two-stage scheme [28] however offers a significant performance gain in terms of outage probability compared to the max-min strategy. In the two-stage relay selection scheme, the first stage ensures the targeted data rate of the users of interest and the second stage opportunistically maximizes the rate of other users.

## 6. Further Opportunities and Challenges

The ultimate goal of employing NOMA is to achieve higher spectral efficiency with reasonable user fairness. It has appeared as a paradigm changer for the design of future radio access techniques. NOMA is compatible to various advanced communication techniques to be used in 5G and Beyond. However, the opportunities come with various challenges, as outlined below:

**NOMA-based Heterogeneous Networks:** In the context of 5G requirements, heterogeneous network (HetNet) is an effective solution to achieve high capacity enhancement, smart utilization of limited resources, and low-cost networks [29]. In HetNets, in addition to primary BS (called macro-BS), there are other types of small BSs such as micro-BS and pico-BS in each cell, as opposed to a homogenous network which has only one BS in each cell. These smaller BSs with lower transmit powers and smaller coverages are installed within the coverage of the macrocell to increase spectral efficiency and the number of served users. However, one of the key challenges in HetNet is to maintain the fairness among users. In order to provide a better user experience, the application of NOMA in HetNet is therefore beneficial. The research that employed massive MIMO in macro cells and NOMA in small cells [30] showed that NOMA-based HetNet outperforms the traditional OMA-based HetNet. NOMA-aided HetNet also significantly improves the downlink coverage performance in cooperative communication scenarios [31]. The NOMA-based two-tier HetNet with non-uniform small cell deployment is found more energy-efficient compared to OMA-enabled HetNet [32]. In NOMA-based multi-cell networks, power control is an essential issue to focus on, because both intra-cell interference and inter-cell interference need to be coordinated. Otherwise, the performance of cell edge users can be severely degraded [33]. Whereas there exists substantial volume of research contributions on both NOMA and HetNet, resource allocation in NOMA-assisted HetNets remains an active area of investigation.

**NOMA in mmWave Communications:** mmWave communication [34] represents one of the most promising technologies for 5G and Beyond because of its huge bandwidth resource. Its applications are feasible in various contexts such as IoT and cloud-assisted vehicular networks [35]. However, the performance of mmWave systems is severely degraded because its transmission is highly directional, and thus, users' channels are highly correlated. In contrast, such high correlation is a favorable condition for the application of NOMA [36]. The integration of NOMA with mmWave is therefore highly appropriate to support massive connectivity in dense networks. In fact, the application of NOMA in mmWave has been investigated in different settings and scenarios, and NOMA is shown to be more efficient than OMA [36-39, 83]. Despite its potentiality, research on NOMA-based mmWave communication is still in the early stage. Therefore, multidimensional studies such as optimization of resource allocation, cooperative mmWave-NOMA networks, and mmWave-NOMA-aided Hetnets are required to further boost the system performance.

**NOMA in Visible Light Communications:** In addition to mmWave communications, another communication paradigm with rich spectrum resource that has attracted a significant attention in recent years is visible light communication (VLC) [40]. Since light is not able to penetrate through walls, VLC inherently offers a higher level of information security. Moreover, VLC is particularly useful in some sensitive environments such as aircraft cabins and hospitals, where the interference to existing radio frequency (RF) systems is by nature problematic. Despite major differences between the VLC channel and the RF channel, the application of NOMA to downlink VLC case has the potential to further increase the performance of VLC networks without affecting light emitting diode (LED) lighting quality [41]. Further research needs to be carried out to make NOMA-enabled VLC workable in different settings and scenarios including NOMA in uplink VLC and MIMO-NOMA in VLC. Resource allocation in NOMA-aided VLC is also an open problem.

**Other Open Issues:** Given the dynamic nature of user demands, systems, networks, services, applications and traffic patterns in 5G and Beyond, there still exist different open issues and research challenges to utilize the ultimate potentials of NOMA. The next generation wireless network needs to accommodate hybrid multiple access, where NOMA will be integrated with other multiple access techniques. Such integration can be done via either single-tone approach or multi-tone approach. In the former, NOMA is implemented on each orthogonal resource element (e.g. an OFDM subcarrier within an assigned time slot) individually. Power domain NOMA is an example of single-tone NOMA, where the implementation of NOMA on one resource element does not necessarily become dependent to those on the others. The latter category of integration jointly implements NOMA across multiple orthogonal resource elements [13, 42], some sorts of joint coding across the multiple resource elements are performed. Conversely, low-complexity optimal resource allocation for multi-tone NOMA is currently a significant research direction. Then again, NOMA can be utilized beyond cellular communications such as digital television broadcasting [43], wireless content caching [44] and underwater communication [84]. However, the NOMA designs should be tailored to meet the requirements of each new application. For example, how the content popularity can be taken into account in designing of a NOMA-assisted content caching is yet to be elucidated.

## 7. Summary

In this article, we present the basic principles of NOMA and discuss how it becomes advantageous for the next generation wireless networks. The article first elaborates some fundamentals of two broad categories of NOMA: power-domain and code-domain, in downlink and uplink environments. Then, the article focuses on power-domain NOMA. Compared to downlink NOMA, uplink NOMA is more appropriate to serve a large number of users because computation-intensive decoding is performed at the BS. Key benefits of using NOMA as a radio access technique for 5G and Beyond are high spectral efficiency, massive connectivity, and low latency. With appropriate power allocation, NOMA offers enhanced user fairness and thus improves cell-edge users' experiences. To understand how MIMO technologies can be integrated with NOMA for further system performance improvement, the article provides an introductory lesson on cluster-based MIMO-NOMA and beamforming-based MIMO-NOMA in downlink setting. Compared to cluster-based MIMO-NOMA, beamforming-based MIMO-NOMA offers improved system performances at the expense of additional implementation complexity. Although research works on MIMO-NOMA in uplink setting is yet to be enriched, we offer some insights into it. Like other wireless communication systems, resource allocation is the central aspect of NOMA. On that, the article pays a fair attention to cover the important concepts of resource allocation including user clustering and power allocation. For the user clustering methods, channel gain-based approaches are heuristic in principle and thus come with low-complexity but unstable system performances. By contrast, various systematic frameworks such as combinatorial relaxation and matching theory-based approaches are promising in terms of tradeoff between system performance and computational complexity. There exist significant differences in the power allocation strategies of SC-NOMA and MC-NOMA because of sub-carrier assignment in the latter. Further, the article discusses the concept of cooperative NOMA that appears in different forms. Each of the variants of cooperative NOMA comes with unique features and offers improved performances compared to the OMA counterpart. Finally, the article focuses on several open opportunities and challenges that must be addressed to make NOMA compatible with other promising communication paradigms such as heterogeneous networks, mmWave communications, and content caching.

**References**

[1] Boccuzzi J. (2019) Introduction to Cellular Mobile Communications. In: Vaezi M., Ding Z., Poor H. (eds) Multiple Access Techniques for 5G Wireless Networks and Beyond. Springer, Cham.

[2] D. Tse and P. Viswanath, Fundamentals of Wireless Communication. Cambridge, U.K.: Cambridge Univ. Press, 2005.

[3] A. Benjebbovu, A. Li, Y. Saito, Y. Kishiyama, A. Harada, and T. Nakamura, "System-level Performance of Downlink NOMA for Future LTE Enhancements," in Proc. IEEE Global Communications Conference (GLOBECOM), Atlanta, USA, 2013, pp. 66–70.

[4] S. M. R. Islam, N. Avazov, O. A. Dobre and K. Kwak, "Power-Domain Non-Orthogonal Multiple Access (NOMA) in 5G Systems: Potentials and Challenges," IEEE Communications Surveys and Tutorials, vol. 19, no. 2, pp. 721-742, Second quarter 2017.

[5] Z. Ding, X. Lei, G. K. Karagiannidis, R. Schober, J. Yuan and V. K. Bhargava, "A Survey on Non-Orthogonal Multiple Access for 5G Networks: Research Challenges and Future Trends," IEEE Journal on Selected Areas in Communications, vol. 35, no. 10, pp. 2181-2195, Oct. 2017.

[6] Y. Liu, Z. Qin, M. Elkashlan, Z. Ding, A. Nallanathan and L. Hanzo, "Nonorthogonal Multiple Access for 5G and Beyond," Proceedings of the IEEE, vol. 105, no. 12, pp. 2347-2381, Dec. 2017.

[7] K. Higuchi and A. Benjebbour, "Non-Orthogonal Multiple Access (NOMA) with Successive Interference Cancellation," IEICE Transactions on Communications, vol. E98-B, no. 3, pp. 403–414, Mar. 2015.

[8] L. Dai, B. Wang, Z. Ding, Z. Wang, S. Chen and L. Hanzo, "A Survey of Non-Orthogonal Multiple Access for 5G," IEEE Communications Surveys & Tutorials, vol. 20, no. 3, pp. 2294-2323, third quarter 2018.

[9] SMR Islam, M Zeng, and Octavia A. Dobre, "NOMA in 5G Systems: Exciting Possibilities for Enhancing Spectral Efficiency," IEEE 5G Tech Foucs, vol. 1, no. 2, Jun. 2017.

[10] M. Mohammadkarimi, M. A. Raza and O. A. Dobre, "Signature-Based Non-orthogonal Massive Multiple Access for Future Wireless Networks: Uplink Massive Connectivity for Machine-Type Communications," IEEE Vehicular Technology Magazine, vol. 13, no. 4, pp. 40-50, Dec. 2018.

[11] R. Hoshyar, F. P. Wathan, and R. Tafazolli, "Novel Low-Density Signature for Synchronous CDMA Systems over AWGN Channel," IEEE Transactions on Signal Processing, vol. 56, no. 4, pp. 1616–1626, Apr. 2008.

[12] R. Hoshyar, R. Razavi, and M. Al-Imari, "LDS-OFDM an Efficient Multiple Access Technique," in Proc. IEEE Vehicular Technology Conference (IEEE VTC Spring), Taipei, Taiwan, 2010, pp. 1–5.

[13] H. Nikopour and H. Baligh, "Sparse Code Multiple Access," in Proc. IEEE International Symposium on Personal Indoor Mobile Radio Communications (IEEE PIMRC), London, U.K., 2013, pp. 332–336.

[14] S. Timotheou and I. Krikidis, "Fairness for Non-Orthogonal Multiple Access in 5G Systems," IEEE Signal Processing Letters, vol. 22, no. 10, pp. 1647–1651, Oct. 2015.

[15] A. Bayesteh, E. Yi, H. Nikopour, and H. Baligh, "Blind Detection of SCMA for Uplink Grant-free Multiple-Access," in Proc. IEEE Wireless Communications Systems, Barcelona, Spain, 2014, pp. 853–857.

[16] H. Tabassum, M. S. Ali, E. Hossain, M. J. Hossain and D. I. Kim, "Uplink Vs. Downlink NOMA in Cellular Networks: Challenges and Research Directions," in Proc. IEEE Vehicular Technology Conference (VTC Spring), Sydney, Australia, 2017, pp. 1-7.

[17] Z. Ding, M. Peng and H. V. Poor, "Cooperative Non-Orthogonal Multiple Access in 5G Systems," IEEE Communications Letters, vol. 19, no. 8, pp. 1462-1465, Aug. 2015.

[18] J. N. Laneman, D. N. C. Tse, and G. W. Wornell, "Cooperative Diversity in Wireless Networks: Efficient Protocols and Outage Behavior," IEEE Transactions on Information Theory, vol. 50, no. 12, pp. 3062–3080, Dec. 2004.

[19] J. B. Kim and I. H. Lee, "Capacity Analysis of Cooperative Relaying Systems Using Non-orthogonal Multiple Access," IEEE Communications Letters, vol. 19, no. 11, pp. 1949-1952, Nov. 2015.

[20] R. Jiao, L. Dai, J. Zhang, R. MacKenzie, and M. Hao, "On the Performance of NOMA-based Cooperative Relaying Systems over Rician Fading Channels," IEEE Transactions on Vehicular Technology, vol. 66, no. 12, pp. 11409-11413, Dec. 2017.

[21] M. Xu, F. Ji, M. Wen and W. Duan, "Novel Receiver Design for the Cooperative Relaying System with Non-orthogonal Multiple Access," IEEE Communications Letters, vol. 20, no. 8, pp. 1679-1682, Aug. 2016.

[22] W. Shin, H. Yang, M. Vaezi, J. Lee and H. V. Poor, "Relay-aided NOMA in Uplink Cellular Networks," IEEE Signal Processing Letters, vol. 24, no. 12, pp. 1842-1846, Dec. 2017.

[23] M. F. Kader, M. B. Shahab and S. Y. Shin, "Exploiting Non-orthogonal Multiple Access in Cooperative Relay Sharing," IEEE Communications Letters, vol. 21, no. 5, pp. 1159-1162, May 2017.

[24] M. F. Kader, M. B. Uddin, SMR Islam and S. Y. Shin, "Capacity and Outage Analysis of a Dual-Hop Decode-and-Forward Relay-Aided NOMA Scheme," Digital Signal Processing, vol. 88, pp. 138-148, May 2019.

[25] J. Kim and I. Lee, "Non-Orthogonal Multiple Access in Coordinated Direct and Relay Transmission," IEEE Communications Letters, vol. 19, no. 11, pp. 2037-2040, Nov. 2015.

[26] J. Men and J. Ge, "Non-orthogonal Multiple Access for Multiple-antenna Relaying Networks", IEEE Communications Letters, vol. 19, no. 10, pp. 1686-1689, Oct. 2015.

[27] Y. Jing and H. Jafarkhani, "Single and Multiple Relay Selection Schemes and Their Achievable Diversity Orders," IEEE Transactions on Wireless Communications, vol. 8, no. 3, pp. 1414–1423, Mar. 2009.

[28] Z. Ding, H. Dai and H. V. Poor, "Relay Selection for Cooperative NOMA," in IEEE Wireless Communications Letters, vol. 5, no. 4, pp. 416-419, Aug. 2016.

[29] S. Yuan and Q. Liang, "Cooperative Bandwidth Sharing for 5G Heterogeneous Network Using Game Theory," in IEEE International Conference on Networking, Architecture and Storage (NAS), Long Beach, CA, 2016, pp. 1-6.

[30] Y. Liu, Z. Qin, M. Elkashlan, A. Nallanathan, J. A. McCann, "Non-Orthogonal Multiple Access in Large-Scale Heterogeneous Networks", IEEE Journal on Selected Areas in Communications, vol. 35, no. 12, pp. 2667-2680, Dec. 2017.

[31] C.-H. Liu, D.-C. Liang, P.-C. Chen and J.-R. Yang, "Coverage Analysis for Dense Heterogeneous Networks with Cooperative NOMA", in Proc. IEEE Vehicular Technology Conference, Toronto, Canada, 2017, pp. 1-6.

[32] T. Han, J. Gong, Z. Lio, SMR Islam, Q. Li, Z. Bai and K. S. Kwak, "On Downlink NOMA in Heterogeneous Networks with Non-Uniform Small Cell Deployment," IEEE Access, vol. 6. pp. 31099-31109, Jun. 2018.

[33] W. Shin, M. Vaezi, B. Lee, D.J. Love, J. Lee and H. V. Poor, "Non-Orthogonal Multiple Access in Multi-Cell Networks: Theory, Performance, and Practical Challenges," IEEE Communications Magazine, vol. 55, no. 10, pp. 176–183, Oct. 2017.

[34] T. S. Rappaport et al., "Millimeter Wave Mobile Communications for 5G Cellular: It Will Work!," IEEE Access, vol. 1, pp. 335-349, May 2013.

[35] L. Kong, M. K. Khan, F. Wu, G. Chen and P. Zeng, "Millimeter-Wave Wireless Communications for IoT-Cloud Supported Autonomous Vehicles: Overview, Design, and Challenges," IEEE Communications Magazine, vol. 55, no. 1, pp. 62–68, Jan. 2017.

[36] Z. Ding, P. Fan and H. V. Poor, "Random Beamforming in Millimeter-Wave NOMA Networks," IEEE Access, vol. 5, pp. 7667-7681, Feb. 2017.

[37] W. Hao, M. Zeng, Z. Chu and S. Yang, "Energy-Efficient Power Allocation in Millimeter Wave Massive MIMO with Non-Orthogonal Multiple Access," IEEE Wireless Communications Letters, vol. 6, no. 6, pp. 782-785, Dec. 2017.

[38] T. Lv, Y. Ma, J. Zeng and P. T. Mathiopoulos, "Millimeter-Wave NOMA Transmission in Cellular M2M Communications for Internet of Things," IEEE Internet of Things Journal, vol. 5, no. 3, pp. 1989-2000, Jun. 2018.

[39] Y. Zhou, V. W. S. Wong and R. Schober, "Coverage and Rate Analysis of Millimeter Wave NOMA Networks with Beam Misalignment," IEEE Transactions on Wireless Communications, vol. 17, no. 12, pp. 8211-8227, Dec. 2018.

[40] S. Arnon, Visible Light Communication. Cambridge University Press, New York, NY, USA, 2015.

[41] L. Yin, W. O. Popoola, X. Wu and H. Haas, "Performance Evaluation of Non-Orthogonal Multiple Access in Visible Light Communication," IEEE Transactions on Communications, vol. 64, no. 12, pp. 5162-5175, Dec. 2016.

[42] Y. Sun, D.W.K. Ng, Z. Ding and R. Schober, "Optimal Joint Power and Subcarrier Allocation for Full-duplex Multicarrier Non-orthogonal Multiple Access Systems," IEEE Transactions on Communications, vol. 65, no. 3, pp. 1077–1091, Mar. 2017.

[43] N. A. K. Beigi and M. R. Soleymani, "Cooperative NOMA in Multi-Content Multimedia Broadcasting," in Proc. IEEE International Conference on Communications Workshops (ICC Workshops), Kansas City, USA, 2018, pp. 1-6.

[44] Z. Ding, P. Fan, G. K. Karagiannidis, R. Schober and H. V. Poor, "NOMA Assisted Wireless Caching: Strategies and Performance Analysis," in IEEE Transactions on Communications, vol. 66, no. 10, pp. 4854-4876, Oct. 2018.

[45] "Study on Downlink Multiuser Superposition Transmission for LTE", Document, 3rd Generation Partnership Project (3GPP), Jun. 2015.

[46] Z. Chen, Z. Ding, P. Xu and X. Dai, "Optimal Precoding for a QoS Optimization Problem in Two-user MISO-NOMA Downlink," IEEE Communications Letters, vol. 20, no. 6, pp. 1263–1266, Jun. 2016.

[47] Z. Ding, F. Adachi and H. V. Poor, "The Application of MIMO to Non-orthogonal Multiple Access," IEEE Transactions on Wireless Communications, vol. 15, no. 1, pp. 537–552, Jan. 2016.

[48] Z. Ding, R. Schober and H. V. Poor, "A General MIMO Framework for NOMA Downlink and Uplink Transmission Based on Signal Alignment," IEEE Transactions on Wireless Communications, vol. 15, no. 6, pp. 4438–4454, Jun. 2016.

[49] M. Zeng, A. Yadav, O. A. Dobre, G. I. Tsiropoulos and H. V. Poor, "Capacity Comparison between MIMO-NOMA and MIMO-OMA with Multiple Users in a Cluster," IEEE Journal on Selected Areas in Communications, vol. 35, no. 10, pp. 2413–2424, Oct. 2017.

[50] —, "On the Sum Rate of MIMO-NOMA and MIMO-OMA Systems," IEEE Wireless Communications letters, vol. 6, no. 4, pp. 534–537, Aug. 2017.

[51] Y. Saito et al., "Non-orthogonal Multiple Access (NOMA) for Cellular Future Radio Access," in Proc. IEEE Vehicular Technology Conference (VTC Spring), Dresden, Germany, Jun. 2013, pp. 1–5.

[52] B. Kimy et al., "Non-orthogonal Multiple Access in a Downlink Multiuser Beamforming System," in Proc. IEEE Military Communications Conference (MILCOM), San Diego, CA, USA, Nov. 2013, pp. 1278–1283.

[53] K. Higuchi and Y. Kishiyama, "Non-orthogonal Access with Random Beamforming and Intra-beam SIC for Cellular MIMO Downlink," in Proc. IEEE Vehicular Technology Conference (VTC Fall), Las Vegas, NV, USA, Sep. 2013, pp. 1–5.

[54] Q. Sun, S. Han, I. Chin-Lin and Z. Pan, "On the Ergodic Capacity of MIMO NOMA systems," IEEE Wireless Communications letters, vol. 4, no. 4, pp. 405–408, Dec. 2015.

[55] J. Choi, "On the Power Allocation for MIMO-NOMA Systems with Layered Transmissions," IEEE Transactions on Wireless Communications, vol. 15, no. 5, pp. 3226–3237, May 2016.

[56] M. F. Hanif, Z. Ding, T. Ratnarajah and G. K. Karagiannidis, "A Minorization-maximization Method for Optimizing Sum Rate in the Downlink of Non-orthogonal Multiple Access Systems," IEEE Transations on Signal Processing, vol. 64, no. 1, pp. 76–88, Jan. 2016.

[57] M. Zeng, W. Hao, O. A. Dobre and H. V. Poor, "Energy-efficient Power Allocation in Uplink mmWave Massive MIMO with NOMA," IEEE Transactions on Vehicular Technology, vol. 68, no. 3, pp. 3000-3004, Mar. 2019.

[58] L. Salaun, C. S. Chen and M. Coupechoux, "Optimal Joint Subcarrier and Power Allocation in NOMA is Strongly NP-hard," in Proc. IEEE International Conference on Communications (ICC), Kansas City, MO, USA, May 2018, pp. 1–7.

[59] L. Lei, D. Yuan, C. K. Ho and S. Sun, "Power and Channel Allocation for Non-orthogonal Multiple Access in 5G Systems: Tractability and Computation," IEEE Transactions on Wireless Communications, vol. 15, no. 12, pp. 8580–8594, Dec. 2016.

[60] B. Di, L. Song and Y. Li, "Sub-channel assignment, Power Allocation, and User Scheduling for Non-orthogonal Multiple Access Networks," IEEE Transactions on Wireless Communications, vol. 15, no. 11, pp. 7686–7698, Nov. 2016.

[61] Z. Ding, P. Fan and H. V. Poor, "Impact of User Pairing on 5G Non-orthogonal Multiple Access Downlink Transmissions," IEEE Transactions on Vehicular Technology, vol. 65, no. 8, pp. 6010–6023, Aug. 2016.

[62] S. M. R. Islam, M. Zeng, O. A. Dobre and K. Kwak, "Resource Allocation for Downlink NOMA Systems: Key Techniques and Open Issues," IEEE Wireless Communications Magazine, vol. 25, no. 2, pp. 40–47, April 2018.

[63] L. Zhu, J. Zhang, Z. Xiao, X. Cao and D. O. Wu, "Optimal User Pairing for Downlink Non-orthogonal Multiple Access (NOMA)," IEEE Wireless Communications letters, vol. 8, no. 2, pp. 328–331, Apr. 2019.

[64] Y. Sun, D. W. K. Ng, Z. Ding and R. Schober, "Optimal Joint Power and Subcarrier Allocation for MC-NOMA systems," in Proc. IEEE Global Communications Conference (GLOBECOM), Washington, DC, USA, Dec 2016, pp. 1–6.

[65] K. Wang, Z. Ding and W. Liang, "A Game Theory Approach for User Grouping in Hybrid Non-orthogonal Multiple Access Systems," in Proc. IEEE International Symposium on Wireless Communication Systems (ISWCS), Poznan, Poland, Sep. 2016, pp. 643–647.

[66] K. Wang, J. Cui, Z. Ding and P. Fan, "A Stackelberg Game Approach for NOMA in mmwave Systems," in Proc. IEEE Global Communications Conference (GLOBECOM), Abu Dhabi, United Arab Emirates, Dec 2018, pp. 1–6.

[67] J. Ding, J. Cai and C. Yi, "An Improved Coalition Game Approach for MIMO-NOMA Clustering Integrating Beamforming and Power Allocation," IEEE Transactions on Vehicular Technology, vol. 68, no. 2, pp. 1672–1687, Feb. 2019.

[68] Y. Gu, W. Saad, M. Bennis, M. Debbah and Z. Han, "Matching Theory for Future Wireless Networks: Fundamentals and Applications," IEEE Communications Magazine, vol. 53, no. 5, pp. 52–59, May 2015.

[69] M. Zeng, A. Yadav, O. A. Dobre and H. V. Poor, "Energy-efficient Joint User-RB Association and Power Allocation for Uplink Hybrid NOMA-OMA," IEEE Internet of Things Journal, pp. 1–1, 2019, doi: 10.1109/JIOT.2019.2896946.

[70] W. Liang, Z. Ding, Y. Li and L. Song, "User Pairing for Downlink Non-orthogonal Multiple Access Networks Using Matching Algorithm," IEEE Transactions on Communications, vol. 65, no. 12, pp. 5319–5332, Dec. 2017.

[71] M. B. Shahab, M. F. Kader and S. Y. Shin, "A Virtual User Pairing Scheme to Optimally Utilize the Spectrum of Unpaired Users in Non-orthogonal Multiple Access," IEEE Signal Processing Letters, vol. 23, no. 12, pp. 1766–1770, Dec. 2016.

[72] M. Zeng, G. I. Tsiropoulos, O. A. Dobre and M. H. Ahmed, "Power Allocation for Cognitive Radio Networks Employing Non-orthogonal Multiple Access," in Proc. IEEE Global Communications Conference (GLOBECOM), Washington DC, USA, Dec. 2016, pp. 1-5.

[73] M. Zeng, A. Yadav, O. A. Dobre and H. V. Poor, "Energy-efficient Power Allocation for MIMO-NOMA with Multiple Users in a Cluster," IEEE Access, vol. 6, pp. 5170–5181, Feb. 2018.

[74] M. Zeng, G. I. Tsiropoulos, A. Yadav, O. A. Dobre and M. H. Ahmed, "A Two-phase Power Allocation Scheme for CRNs Employing NOMA," in Proc. IEEE Global Communications Conference (GLOBECOM), Singapore, Singapore, Dec. 2017, pp. 1–6.

[75] M. Zeng, A. Yadav, O. A. Dobre and H. V. Poor, "Energy-efficient Power Allocation for Uplink NOMA," in Proc IEEE Global Communications Conference (GLOBECOM), Abu Dhabi, UAE, Dec. 2018, pp. 1–6.

[76] M. Zeng, N. P. Nguyen, O. A. Dobre and H. V. Poor, "Securing Downlink Massive MIMO-NOMA Networks with Artificial Noise," IEEE Journals on Selected Topics in Signal Processing, pp. 1–1, Feb. 2019, doi: 10.1109/JSTSP.2019.2901170.

[77] S. Shi, L. Yang and H. Zhu, "Outage Balancing in Downlink Nonorthogonal Multiple Access with Statistical Channel State Information," IEEE Transactions on Wireless Communications, vol. 15, no. 7, pp. 4718–4731, Jul. 2016.

[78] M. Zeng, A. Yadav, O. A. Dobre and H. V. Poor, "A Fair Individual Rate Comparison between MIMO-NOMA and MIMO-OMA," in Proc IEEE Global Communications Conference (GLOBECOM) Wkshps, Singapore, Dec 2017, pp. 1–5.

[79] —, "Energy-efficient Power Allocation for Hybrid Multiple Access Systems," in Proc. IEEE International Conference on Communications Workshops (ICC Workshops), Kansas City, MO, USA, May 2018, pp. 1–5.

[80] M. Zeng, N. P. Nguyen, O. A. Dobre, Z. Ding and H. V. Poor, "Spectral and Energy Efficient Resource Allocation for Multi-carrier Uplink NOMA Systems," IEEE Transactions on Vehicular Technology, submitted.

[81] Z. Wei, D. W. K. Ng and J. Yuan, "Power-efficient Resource Allocation for MC-NOMA with Statistical Channel State Information," in Proc. IEEE Global Communications Conference (GLOBECOM), Washington, DC, USA, Dec. 2016, pp. 1–7.

[82] G. I. Tsiropoulos, A. Yadav, M. Zeng and O. A. Dobre, "Cooperation in 5G HetNets: Advanced spectrum access and D2D assisted communications," IEEE Wireless Communications Magazine, vol. 24, no. 5, pp. 110–117, Oct. 2017.

[83] W. Hao, Z. Chu, F. Zhou, S. Yang, G. Sun and K. Wong, "Green Communication for NOMA-Based CRAN," in IEEE Internet of Things Journal, vol. 6, no. 1, pp. 666-678, Feb. 2019.

[84] E. Makled, A. Yadav, O. A. Dobre, and R. Haynes, "Hierarchical Full-duplex Underwater Acoustic Network: A NOMA Approach," in Proc. IEEE Oceans 2018, Charleston, USA, pp. 1-6.